\documentclass[10pt,twocolumn,letterpaper]{article}

\usepackage[pagenumbers]{cvpr} 

\definecolor{cvprblue}{rgb}{0.21,0.49,0.74}
\usepackage[pagebackref,breaklinks,colorlinks,allcolors=cvprblue]{hyperref}
\usepackage{algorithmic}
\usepackage[ruled]{algorithm2e}
\def\paperID{18450} 
\def\confName{CVPR}
\def\confYear{2025}

\title{Exploring the Robustness and Transferability of Patch-Based Adversarial Attacks in Quantized Neural Networks}

\author{Amira Guesmi\\
NYU Abu Dhabi\\
UAE\\
\and
Bassem Ouni\\
Technology Innovation Institute\\
UAE\\
\and
Muhammad Shafique\\
NYU Abu Dhabi\\
UAE\\
}

\begin{document}
\maketitle
\begin{abstract}
Quantized neural networks (QNNs) are increasingly used for efficient deployment of deep learning models on resource-constrained platforms, such as mobile devices and edge computing systems. While quantization reduces model size and computational demands, its impact on adversarial robustness—especially against patch-based attacks—remains inadequately addressed. Patch-based attacks, characterized by localized, high-visibility perturbations, pose significant security risks due to their transferability and resilience. In this study, we systematically evaluate the vulnerability of QNNs to patch-based adversarial attacks across various quantization levels and architectures, focusing on factors that contribute to the robustness of these attacks. 
Contrary to the expectation that quantization might enhance adversarial defenses, our results show that QNNs remain highly susceptible to patch attacks due to the persistence of distinct, localized features within quantized representations. To address this vulnerability, we propose Quantization-Aware Defense Training (QADT), a novel defense strategy that enhances QNN resilience by integrating adversarial patch augmentation and simulated quantized conditions during training. Our results demonstrate that QADT significantly improves the robustness of quantized models against patch-based attacks, reducing attack success rates by a substantial margin. These findings underscore the need for quantization-aware defenses that address the specific challenges posed by patch-based attacks and contribute to a deeper understanding of adversarial robustness in QNNs, guiding future research in developing secure, quantization-compatible defenses for real-world applications.

\end{abstract}    
\section{Introduction}
\label{sec:intro}
As deep learning models are increasingly deployed in resource-constrained environments, such as mobile devices and edge computing platforms, quantization has emerged as a critical method to reduce model size and computational load. By lowering the bitwidth of the weights and activations, quantization enables models to operate more efficiently, preserving accuracy while significantly reducing resource demands \cite{nagel2021white}. QNNs are now central to applications ranging from autonomous systems \cite{liu2021flexi, katare2023survey} to healthcare \cite{zhang2021medq} and Internet of Things (IoT) devices \cite{tonellotto2021neural, hernandez2024optimizing, fedQNN}. However, as QNNs gain prominence, ensuring their robustness against adversarial attacks, a long-standing vulnerability in deep learning, remains an essential concern for secure deployment.

Among adversarial threats, patch-based attacks pose a unique and formidable challenge \cite{guesmi2023physical}. Unlike conventional adversarial perturbations that rely on subtle, dispersed changes, patch attacks introduce highly visible, localized perturbations that disrupt model predictions by “hijacking” the model’s focus. These attacks are especially problematic because of their transferability and resilience across diverse configurations, making them highly effective in physically realizable real-world scenarios. The distinct nature of patch-based attacks raises pressing questions about the robustness of QNNs against these localized, high-intensity adversarial patterns.

Quantization is often viewed as a potential defense against adversarial attacks, with evidence suggesting that reduced precision may mitigate some types of adversarial perturbations \cite{bernhard2019impact, sen2020empir, fu2021double}. However, the effectiveness of quantization in countering patch-based attacks remains largely untested. Given that patch attacks rely on distinct and highly visible features to exploit model vulnerabilities, it is reasonable to suspect that they may retain their efficacy even in quantized models. Most existing studies on adversarial robustness in QNNs have focused on dispersed or pixel-level attacks, overlooking the unique challenges presented by patch-based attacks.

This paper aims to address this gap by systematically investigating the transferability and effectiveness of patch-based attacks across QNNs with different architectures and bitwidths. Through a series of experiments, we examine factors that may contribute to the resilience of patch-based attacks in quantized settings, including quantization shift, gradient alignment, spatial sensitivity, and feature retention. 

\textbf{Our contributions can be summarized as follows:}
\begin{itemize}
    \item We systematically evaluate how patch-based attacks affect feature representations at different quantization levels. Our findings reveal that quantization fails to effectively disrupt the adversarial signal of patches, as their high-visibility features persist across bit widths. This persistence underscores a critical vulnerability in QNNs.
    \item Our experiments demonstrate that patch-based adversarial attacks achieve remarkable transferability across a variety of quantization bit widths and neural network architectures. This highlights the cross-model and cross-quantization robustness of patch attacks, presenting a significant security concern for QNN deployments.
    \item Through a comparative analysis of gradient alignment for patch-based versus pixel-level attacks, we show that patches retain stronger gradient alignment across bit widths. This robustness allows patches to maintain their adversarial effect despite quantization-induced gradient shifts, unlike pixel-level attacks, which degrade significantly under similar conditions.
    \item We investigate the spatial sensitivity of patch attacks by testing shifted patch placements across different locations and quantization levels. Our results reveal that even significant spatial adjustments retain high attack success rates, indicating that patch attacks exploit localized features that are inherently resilient to spatial shifts. 
    \item Based on our findings, we discuss key strategies to enhance the robustness of QNNs against patch-based adversarial attacks and propose Quantization-Aware Defense Training (QADT), a defense to increase QNN resilience to patch-based attacks. 
\end{itemize}

\section{Related Work}
\label{sec:related_work}

QNNs are widely adopted to reduce model size and computational costs, making them suitable for mobile devices and edge applications \cite{nagel2021white}. Quantization-Aware Training (QAT) \cite{hubara2016binarized, rastegari2016xnor, zhou2016dorefa, li2019additive} and Post-Training Quantization (PTQ) \cite{nagel2020up, li2021brecq, hubara2021accurate, wei2022qdrop} are two prominent approaches for implementing low-bitwidth models, each introducing varying degrees of robustness against adversarial attacks \cite{bernhard2019impact, sen2020empir, fu2021double}. Recent studies indicate that while quantization may disrupt the gradient landscape, creating potential resistance to some pixel-level perturbations \cite{goodfellow2014explaining, madry2017towards, carlini2017towards}, its effectiveness against more structured adversarial patterns remains uncertain \cite{zhang2021medq}. Research on QNNs has focused mainly on dispersed, pixel-level adversarial attacks \cite{Li2024InvestigatingTI, yang2024quantization}, leaving a gap in understanding how QNNs respond to localized, high-visibility attacks like patch-based adversarial patterns.

\section{Methodology}
\label{methodology}

\begin{figure*}
    \centering
    \includegraphics[width=0.85\linewidth]{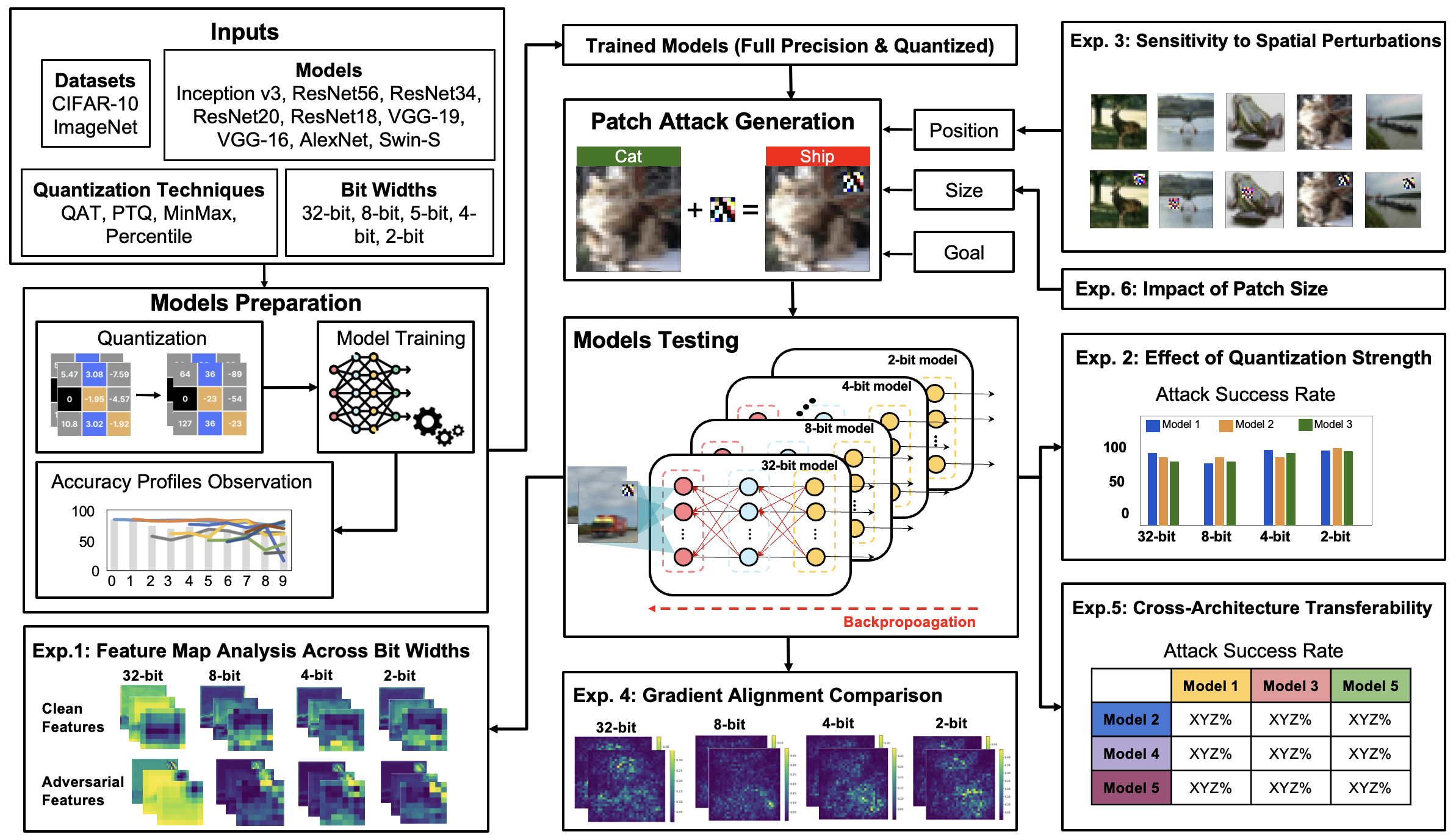}
    \caption{Overview of the experimental framework for evaluating patch-based attacks on QNNs.}
    \label{fig:framework}
\end{figure*}

In this study, we systematically investigate the robustness and transferability of patch-based adversarial attacks across quantized neural networks with different architectures and bitwidths. As illustrated in Figure \ref{fig:framework}, our methodology consists of a series of controlled experiments designed to examine how factors such as quantization level, spatial placement, gradient alignment, and patch size contribute to the success of patch-based attacks in QNNs. 

\subsection{Feature Map Analysis Across Bit Widths:} 
This experiment investigates whether patch-based attacks generate distinct localized feature activations that persist across different bit widths. By analyzing these activations, we aim to understand why patches are resilient to quantization noise and if the model’s focus on the adversarial patch is preserved despite reduced precision. If patches indeed produce prominent activations consistently across bit widths, it would reveal a critical vulnerability in quantized models, suggesting that localized adversarial signals remain impactful even under heavy quantization.\\
\textbf{Hypothesis 1:} If the patch attack is effective due to the creation of distinct, localized features, we expect to observe consistent and prominent patch-induced activations across bit widths and quantization strategies. This would suggest that quantization does not significantly disrupt the model’s focus on the adversarial patch, indicating a persistent vulnerability.
\subsection{Effect of Quantization Levels on Patch Transferability:} 
This experiment evaluates whether the strength of quantization (i.e., reducing bit width) affects the transferability and effectiveness of patch-based attacks. By examining attack success rates across different bit widths, we aim to determine if heavy quantization can disrupt the adversarial impact of patches. High attack success rates at lower bit widths would indicate that patches retain their effectiveness despite quantization, suggesting that quantized models may lack effective defenses against this type of attack.\\
\textbf{Hypothesis 2:} If attack success rates remain high even at lower bit widths, this would suggest that the patch’s strong, localized features continue to influence model behavior despite the added quantization noise, indicating a persistent vulnerability in quantized models.

\subsection{Spatial Sensitivity of Patches:} 
This experiment delves into whether patch attack success relies on precise spatial positioning, a factor especially relevant in the intricate tapestry of quantized models, where spatial representations might subtly shift. If minor changes in patch placement lead to a substantial drop in success rates, it would reveal a dependence on exact spatial orientation. Such findings would suggest that patches exploit specific, localized features that remain effective within the labyrinth of quantized representations.\\
\textbf{Hypothesis 3:} If small spatial adjustments result in a marked decrease in attack success, this would suggest that quantized models are sensitive to precise patch positioning. Such sensitivity would indicate that the patch’s effectiveness relies on spatially localized features that align with the quantized model's internal representations, pointing to a significant spatial dependency in the model’s vulnerability to patch-based attacks.
\vspace{-3mm}
\subsection{Gradient Alignment Across Bit Widths for Patch vs. Pixel-Level Attacks:} 
Analyzing gradient alignment between patch-based and pixel-level attacks can provide insight into why patch attacks exhibit high transferability in quantized models. Greater alignment for patches would suggest that their effectiveness is less impacted by the gradient alterations introduced by quantization, unlike pixel-level perturbations, which depend heavily on precise gradient information and thus experience a decline in effectiveness within quantized models.\\
\textbf{Hypothesis 4:} If gradient alignment for patch attacks remains high across bit widths, it would suggest that patches are less affected by quantization-induced gradient misalignment, contributing to their robustness in quantized models.
\vspace{-3mm}
\subsection{Transferability of Patch Across Architectures:} 
This experiment tests whether adversarial patches generated on one model architecture transfer effectively to other architectures with different quantization levels. By assessing cross-architecture transferability, we aim to understand if patch-based attacks exploit architecture-agnostic features that make them resilient across various quantized models.\\
\textbf{Hypothesis 5:} If the patches retain high success rates across various architectures and bitwidths, this would indicate that they exploit generalized feature vulnerabilities in quantized models. Conversely, a significant drop in success rates would suggest that patch attacks rely on architecture-specific features, making them less effective when transferred to different model structures.
\vspace{-3mm}
\subsection{Impact of Patch Size and Visibility:} 
This experiment examines if the perceptual visibility of patches (e.g., size) correlates with attack success in quantized models. A correlation would suggest that patches rely on their high visibility and distinctiveness, rather than subtle adversarial signals, to remain effective in quantized settings.\\
\textbf{Hypothesis 6:} If larger and more visible patches consistently yield higher success rates, it implies that the strong visibility of the patch, rather than subtle adversarial features, plays a major role in bypassing quantization.

\section{Results and Analysis}
\label{results}
\subsection{Experimental Setup}
\noindent\textbf{Dataset:}
We use the CIFAR-10 dataset \cite{cifar} and the ImageNet dataset \cite{krizhevsky2017imagenet}. 

\noindent\textbf{Model Architectures:}
Our experiments cover a range of architectures including ResNet56, 34, 20, 18 \cite{he2016deep}, VGG-19, 16 \cite{simonyan2014very}, AlexNet \cite{krizhevsky2017imagenet}, Inception-v3 \cite{szegedy2016rethinking}, and the vision transformer Swin-S \cite{liu2021swin}.

\noindent\textbf{Quantization Techniques:}
We use different quantization methods: Quantization-Aware Training and Post-Training Quantization to examine model robustness at varied bitwidths (32-bit, 8-bit, 5-bit, 4-bit, and 2-bit). We also use different calibration techniques such as MinMax and Percentile.

\noindent\textbf{Patch-based Attacks:}
We employ three widely recognized patch-based attack techniques: \textit{LAVAN} \cite{lavan}, \textit{Adversarial Patch GoogleAP (GAP)} \cite{googleap} and \textit{Deformable Patch Representation (DPR)} attack \cite{chen2022shape}. 
Further details are presented in the supplementary material.
\subsection{Experiment 1: Feature Map Analysis Across Bit Widths} 
We evaluate models quantized to different bit widths (32-bit, 8-bit, 4-bit, 2-bit) using QAT. Clean and patched inputs are passed through these models and feature maps are extracted from key intermediate layers. These maps are then visualized to identify differences in representation and persistence of patch-induced features.

\begin{figure*}
    \centering
    \includegraphics[width=0.8\linewidth]{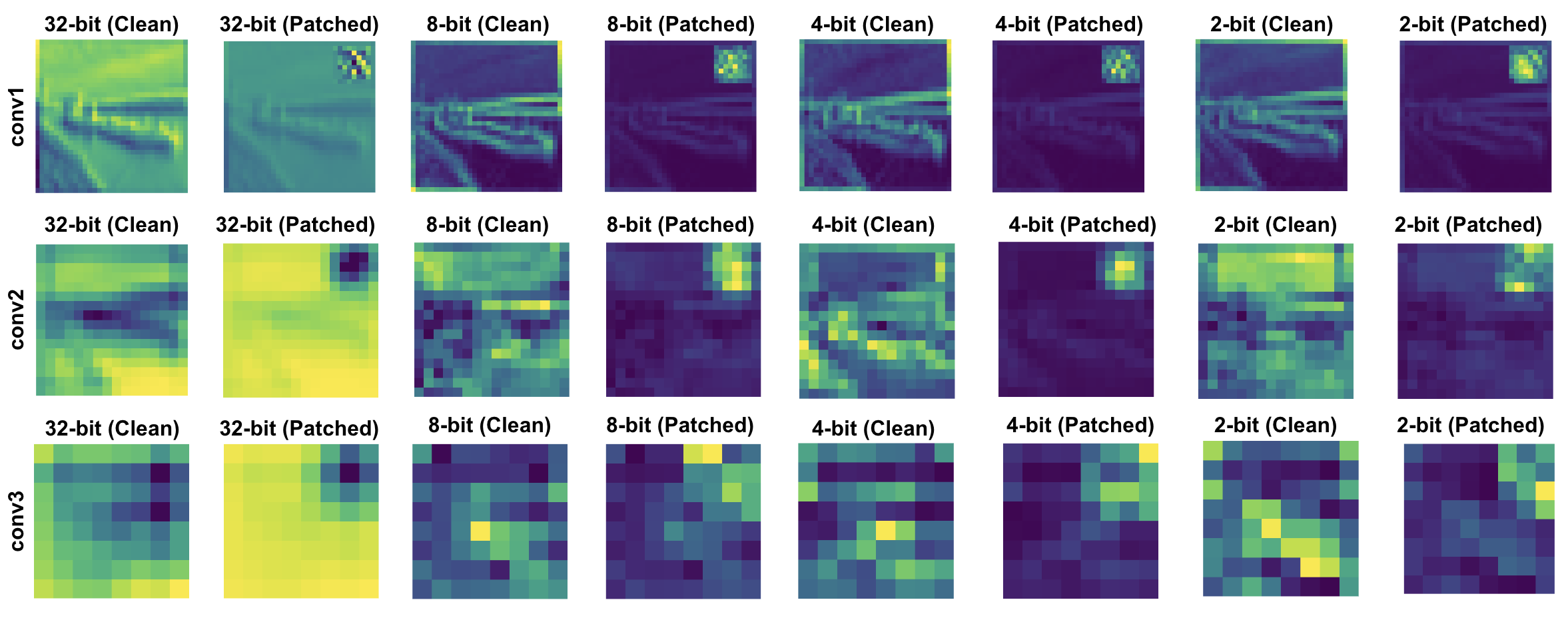}
    \caption{Feature maps of the 32-bit, 8-bit, 4-bit, and 2-bit models comparing the clean and patched feature maps for the three first convolutional layers.}
    \label{fig:experiment1}
\end{figure*}

The feature maps in Figure \ref{fig:experiment1} for patched inputs reveal distinct high-activation regions at the patch location across all bitwidths. In each examined layer, the presence of the patch generates prominent localized activations, which remain visible even at the lowest bitwidths. This observation suggests that the patch induces a strong, persistent signal that is only partially attenuated by quantization.


The persistence of patch-induced activations across all bitwidths indicates that the adversarial signal generated by the patch is resilient to quantization. This robustness is due to the high-visibility patterns created by the patch, which remain effective even at lower precision levels. 
These findings are consistent with the results of the next experiment showing high attack success rates for patch-based attacks across varying quantization levels, suggesting that the patch leverages strong, localized feature disruptions that quantization alone cannot effectively neutralize.

\subsection{Experiment 2: Effect of Quantization Levels on Patch Transferability} 
In this experiment, we generate patch-based attacks using a 32-bit model and then evaluate these patches on the same model quantized to various bit widths (e.g., 8-bit, 4-bit, 2-bit) using both QAT and PTQ. The attack success rate is measured at each bit width to observe any reduction in effectiveness as the bit width decreases.

As shown in Table \ref{tab:experiment2_lavan}, attack success rates remain consistently high across all bitwidths, with mean success rates exceeding 70\% even at the lowest bitwidth (2-bit). This resilience across quantization levels suggests that patch-based attacks maintain their adversarial impact despite a significant reduction in precision, indicating that the patch effect is robust to quantization.

\begin{table}[ht]
    \centering
    \footnotesize  
    \renewcommand{\arraystretch}{0.9}  
    \setlength{\tabcolsep}{3pt}  
    \begin{tabular}{|c|c|c|c|c|c|}
    \hline 
      \textbf{Model} &    \textbf{32-bit} & \textbf{8-bit} & \textbf{5-bit} & \textbf{4-bit} & \textbf{2-bit}\\
    \hline
      ResNet-20  & 87.22  & 83.73  & 80.65 & 77.30 & 74.18 \\ \hline
      ResNet-56  & 86.43  & 83.24  & 79.94 & 76.22 & 73.08 \\ \hline
      VGG-19     & 88.95  & 85.56  & 82.28 & 79.81 & 77.19 \\ \hline
      VGG-16     & 87.17  & 84.73  & 81.45 & 78.29 & 76.67 \\ \hline
      AlexNet    & 82.92  & 79.72  & 77.31 & 75.88 & 73.54 \\ 
    \hline
    \end{tabular}
    \caption{ASR (\%) of LAVAN attack (6x6 patch) across different QNNs on CIFAR-10.}
    \label{tab:experiment2_lavan}
\end{table}
Although attack success rates do decline slightly as bitwidth decreases, the reduction is generally modest. For example, ResNet-20 shows a decrease from 87.22\% at 32-bit to 74.18\% at 2-bit. This pattern suggests that while quantization weakens the patch’s influence to some extent, it is insufficient to neutralize the attack entirely. The high visibility of the patch and distinct features continue to drive effective adversarial responses even under low-precision constraints.

We also evaluate the transferability of patch-based attacks on dynamically quantized models (8-bit) (Table \ref{tab:dyn_ptq}). The results indicate that adversarial patches remain highly effective, confirming that vulnerabilities persist even under dynamic quantization.
\begin{table}[ht]
    \centering
    \footnotesize  
    \renewcommand{\arraystretch}{0.9}  
    \setlength{\tabcolsep}{4pt}  
    \begin{tabular}{|c|c|c|c|c|}
        \hline
        \textbf{Model} & \textbf{ResNet-56} & \textbf{ResNet-20} & \textbf{VGG-19} & \textbf{VGG-16} \\ \hline
        LAVAN & 84.03 & 83.29 & 76.33 & 71.58 \\ \hline
        GAP   & 82.40 & 53.76 & 54.12 & 41.78 \\ \hline
    \end{tabular}
    \caption{ASR (\%) of LAVAN and GAP attacks (6x6 patches) across dynamically quantized models (8-bit) on CIFAR-10.}
    \label{tab:dyn_ptq}
\end{table}

\begin{table}[ht]
    \centering
    \footnotesize  
    \renewcommand{\arraystretch}{0.9}  
    \setlength{\tabcolsep}{3pt}  
    \begin{tabular}{|c|c|c|c|c||c|c|c|c|}
        \hline
         & \multicolumn{4}{c||}{\textbf{ResNet34 (QAT)}} & \multicolumn{4}{c|}{\textbf{ResNet18 (QAT)}} \\ 
        \cline{2-9}
        \textbf{NP} & \textbf{32-bit} & \textbf{5-bit} &  \textbf{4-bit} &\textbf{2-bit} & \textbf{32-bit}  & \textbf{5-bit} & \textbf{4-bit} & \textbf{2-bit} \\ \hline
        0.1 & 99.31 & 66.32 & 63.56 & 56.31 & 99.98 &  72.63 & 67.89 & 65.76 \\ \hline
        0.08 & 98.08 & 64.91 & 59.97  & 52.25 & 99.93 &  66.37& 61.11 & 55.42 \\ \hline
        0.06 & 97.12 & 64.79 & 57.31  & 50.43 & 96.01 & 58.20 & 53.59 & 51.84 \\ \hline
    \end{tabular}
    \caption{ASR (\%) of LAVAN attack with different noise percentages (NP) across different QNNs on ImageNet.}
    \label{tab:experiment2_lavan_resnet34}
\end{table}

\begin{table}[ht]
    \centering
    \footnotesize  
    \renewcommand{\arraystretch}{0.9} 
    \setlength{\tabcolsep}{4pt} 
    \begin{tabular}{|c|c|c|c|c|c|}
        \hline
        \textbf{Model} & \textbf{Quantization} & \textbf{32-bit} & \textbf{8-bit} & \textbf{4-bit} & \textbf{2-bit} \\ \hline
        Inception v3 & QAT & 89.10 & 60.12  & 55.32 & 50.66 \\ \hline
        Swin-S & MinMax & 91.80 & 62.11 & - & - \\ \hline
        Swin-S & Percentile & 93.10 & 63.72 & - & - \\ \hline
    \end{tabular}
    \caption{ASR (\%) of LAVAN attack with a noise percentage 0.08 on ImageNet.}
    \label{tab:experiment2_transformer}
\end{table}

We extended our experiments to ImageNet using ResNet34, ResNet18, Inception-v3, all trained with QAT, along with Swin-S models quantized using PTQ with MinMax and Percentile calibration \cite{LinZSLZ22}.
Quantization fails to efficiently disrupt patch-based attacks (Tables \ref{tab:experiment2_lavan_resnet34} and \ref{tab:experiment2_transformer} )

We evaluated another patch-based attack, the Deformable Patch Representation (DPR) attack \cite{chen2022shape}.
As shown in Table \ref{tab:experiment2_DPR_resnet34}, DPR remains highly effective even on quantized ResNet34, achieving a 49.87\% ASR at 2-bit.

\begin{table}[ht]
    \centering
    \footnotesize  
    \renewcommand{\arraystretch}{0.9}  
    \setlength{\tabcolsep}{4pt}  
    \begin{tabular}{|c|c|c|c|c|}
        \hline
        \textbf{Attack}  & \textbf{32-bit} & \textbf{8-bit} & \textbf{4-bit} & \textbf{2-bit} \\ \hline
        DPR    & 83.46 & 59.32 & 55.82 & 49.87 \\ \hline
    \end{tabular}
    \caption{ASR (\%) of DPR attack against ResNet34 on ImageNet.}
    \label{tab:experiment2_DPR_resnet34}
\end{table}

We also evaluate a more challenging scenario by generating targeted patches, observing that these patches remain transferable across different bitwidths and architectures. Detailed results for this setting are provided in the supplementary material.


\subsection{Experiment 3: Spatial Sensitivity of Patches} 
We generate patches on a 32-bit model and evaluate their effectiveness on QAT models quantized to 8, 5, 4, and 2 bits. For each test, we apply slight spatial modifications to the patch, including shifting by a few pixels, rotating, and scaling. We then record the attack success rate for each modification to assess the impact of these spatial adjustments on model vulnerability.
\begin{table}[ht]
    \centering
    \footnotesize  
    \renewcommand{\arraystretch}{0.9}  
    \setlength{\tabcolsep}{4pt}  
    \begin{tabular}{|c|c|c|c|c|c|}
    \hline
         &  \textbf{32-bit} & \textbf{8-bit} & \textbf{5-bit} & \textbf{4-bit} & \textbf{2-bit}\\
    \hline
      Original            & 86.43  & 83.24  & 79.94 & 76.22 & 73.08 \\ \hline
      Shifted (2,2)       & 87.73  & 82.35  & 79.55 & 77.39 & 72.93 \\ \hline
      Shifted (8,8)       & 85.13  & 82.56  & 80.24 & 78.11 & 72.88 \\ \hline
      Shifted (18,18)     & 82.47  & 80.77  & 79.04 & 74.78 & 70.10 \\ \hline
      Rotated $5^\circ$   & 86.39  & 82.98  &  80.81 & 76.16 & 72.84 \\ \hline
      Rotated $10^\circ$  & 86.24  & 82.68  &  80.45 & 76.79  & 72.20 \\ \hline
      Rotated $20^\circ$  & 84.38  & 81.50  &  79.85 & 75.17  & 71.73 \\ \hline
      Rotated $30^\circ$  & 82.98  & 80.71  &  78.19 & 74.82 & 70.39 \\
    \hline
    \end{tabular}
    \caption{ASR (\%) of LAVAN attack against ResNet-56 (full precision and quantized) on CIFAR-10 at various patch positions and rotation angle. }
    \label{tab:experiment3}
\end{table}


As shown in Table \ref{tab:experiment3}, the original patch maintains high success rates across all bitwidths, from 86.43\% on the 32-bit model to 73.08\% on the 2-bit model. Minor shifts in patch position, such as (2,2) and (8,8), do not significantly impact attack effectiveness, with success rates remaining comparable to those of the original patch across quantized versions. 



Patches maintain high attack success rates across all tested rotation angles, demonstrating resilience to spatial orientation changes. This robustness persists across different quantization levels, although success rates decline slightly at lower bit widths (e.g., 2-bit).  


\subsection{Experiment 4: Gradient Alignment Across Bit Widths for Patch vs. Pixel-Level Attacks} 
We generate gradients for patch-based and pixel-level attacks on a 32-bit model and then measure the cosine similarity between these gradients and those on 8-bit, 4-bit, and 2-bit QAT models.

\begin{figure*}
    \centering
    \includegraphics[width=0.9\linewidth]{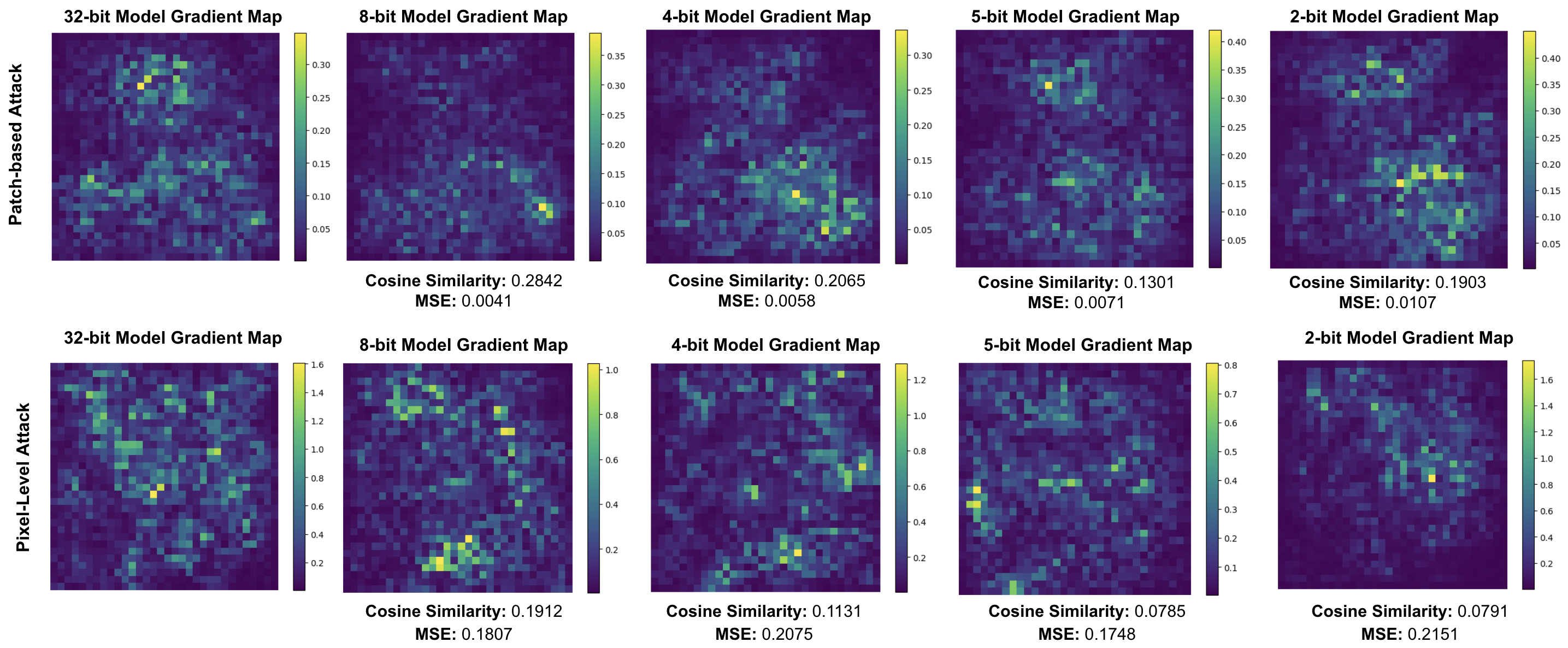}
    \caption{Gradient maps for 32-bit, 8-bit, 4-bit, and 2-bit models under patch-based and pixel-level attacks, along with Cosine Similarity and MSE measurements comparing gradients between the full-precision and quantized models.}
    \label{fig:experiment4}
\end{figure*}

Figure \ref{fig:experiment4} presents gradient maps for both patch-based and pixel-level attacks (PGD) across different quantization levels. For patch-based attacks, the cosine similarity between gradients of the 32-bit model and the quantized models remains relatively high across bitwidths compared to pixel-level attacks. Specifically, the cosine similarity for patch-based attacks decreases gradually from 0.2842 at 8-bit to 0.1301 at 5-bit, consistently surpassing that of pixel-level attacks, which drops as low as 0.0791 at 2-bit. This suggests that patch-based attacks retain a higher degree of gradient alignment across bitwidths, while pixel-level attacks experience more severe gradient misalignment as a result of quantization.

The MSE values for patch-based attacks are consistently lower than those for pixel-level attacks across all bitwidths. For instance, the MSE for patch-based attacks begins at 0.0041 (8-bit) and reaches 0.0107 (2-bit), whereas pixel-level attacks show significantly higher MSE values, starting at 0.1807 (8-bit) and rising to 0.2151 (2-bit). The lower MSE values for patch-based attacks indicate that their gradient magnitudes remain closer to those of the full-precision model, suggesting that quantization introduces less disruption to the gradient structure of patch-based attacks.

In the gradient maps for patch-based attacks, certain high-activation regions remain consistent across bitwidths, preserving visually similar patterns even under increased quantization. In contrast, gradients for pixel-level attacks become increasingly unstructured and diffuse as bitwidth decreases, reflecting a loss of fine-grained detail that is significantly disrupted by quantization.

The higher cosine similarity and lower MSE values observed for patch-based attacks suggest that these attacks retain stronger alignment with the full-precision model’s gradients, even as bitwidth decreases. This robustness indicates that patch attacks leverage feature patterns that quantization does not sufficiently disrupt, likely due to the highly localized and visually distinct characteristics of the patch. This finding aligns with prior observations that patch-based attacks remain effective across quantized models, as the adversarial signal generated by the patch preserves its structure and gradient alignment better than pixel-level perturbations.

In contrast, the significant drop in cosine similarity and increase in MSE for pixel-level attacks across bitwidths demonstrate that quantization disrupts these attacks more effectively. Pixel-level perturbations, which rely on subtle and distributed gradient shifts, are more vulnerable to the gradient misalignment introduced by quantization, resulting in reduced attack robustness in quantized models. 
\subsection{Experiment 5: Transferability of Patches Across Architectures} 
We generate adversarial patches on a base architecture (e.g., Resnet-20 at 32-bit) with high attack success. We then transfer these patches to models with different architectures (e.g., Resnet-56, VGG-16, VGG-19) quantized using QAT at bitwidths such as 8-bit, 5-bit, 4-bit, and 2-bit. We measure and record the attack success rate across architectures and bitwidths.

As shown in Table \ref{tab:experimet5}, attack success rates remain consistently high across different architectures, even when patches are transferred between models. For example, a patch generated on ResNet-20 achieves an 84.17\% success rate on 32-bit ResNet-56 and 78.82\% on 32-bit VGG-19. Similarly, patches created on VGG-19 and VGG-16 maintain high success rates when tested on ResNet architectures, indicating strong cross-architecture transferability of patch-based attacks. This suggests that patch-based attacks are not heavily reliant on architecture-specific features, likely due to their highly visible and localized impact.

Quantization appears to have only a minor impact on the effectiveness of the transferred patches. For instance, when transferring a patch from ResNet-20 to ResNet-56, success rates remain high even at 2-bit (75.21\%). Similarly, transferring a patch from ResNet-56 to VGG-19 yields a high success rate of 71.22\% at 2-bit. This minimal drop in effectiveness across bitwidths suggests that quantization does not significantly hinder the transferability of patch-based attacks, highlighting the resilience of these attacks to precision reduction.
Additional results are provided in the supplementary material.
\begin{table}[ht]
    \centering
    \footnotesize  
    \renewcommand{\arraystretch}{0.9}  
    \setlength{\tabcolsep}{4pt}  
    \begin{tabular}{|c|c|c|c|c|c|c|c|c|c|}
    \hline
          &   \multicolumn{4}{c|}{ \textbf{Res.56}} &  \multicolumn{4}{c|}{\textbf{VGG-19}}  \\ \hline
          & \textbf{32-bit} & \textbf{8-bit}& \textbf{4-bit} & \textbf{2-bit}& \textbf{32-bit} & \textbf{8-bit}  & \textbf{4-bit} & \textbf{2-bit}\\
    \hline                 
      Res.20 &     84.17 & 79.62  & 77.66 & 75.21  &  78.82 & 74.53 & 72.15 & 70.21 \\
    \hline
          & \multicolumn{4}{c|}{ \textbf{Res.20} }& \multicolumn{4}{c|}{  \textbf{VGG-19}}  \\ \hline
          & \textbf{32-bit} & \textbf{8-bit} & \textbf{4-bit} & \textbf{2-bit}& \textbf{32-bit} & \textbf{8-bit} & \textbf{4-bit} & \textbf{2-bit}\\
          \hline      
      Res.56 &  84.11  & 77.67  & 75.33 & 71.76  & 77.43 & 75.09  & 73.82 & 71.22\\
    \hline
           & \multicolumn{4}{c|}{  \textbf{Res.20} }& \multicolumn{4}{c|}{  \textbf{VGG-16}} \\ \hline
          & \textbf{32-bit} & \textbf{8-bit} & \textbf{4-bit} & \textbf{2-bit}& \textbf{32-bit} & \textbf{8-bit}  & \textbf{4-bit} & \textbf{2-bit}\\
          \hline     
      VGG19 &  83.23  & 80.87  & 78.11 &  75.32 & 85.32 & 80.42  & 78.23 & 76.44\\
    \hline
            & \multicolumn{4}{c|}{ \textbf{VGG-19}} &  \multicolumn{4}{c|}{  \textbf{Res.56} } \\ \hline
          & \textbf{32-bit} & \textbf{8-bit} & \textbf{4-bit} & \textbf{2-bit}& \textbf{32-bit} & \textbf{8-bit}  & \textbf{4-bit} & \textbf{2-bit}\\
          \hline     
      VGG16  & 83.29 & 78.48 & 76.65 &  74.39 & 80.55 &  78.93  & 75.34  & 73.87 \\
    \hline
    \end{tabular}
    \caption{ASR (\%) transfer across different QNNs with different bitwidths and architectures on CIFAR-10. }
    \label{tab:experimet5}
\end{table}




\subsection{Experiment 6: Impact of Patch Size and Visibility}  %
We generate patch-based attacks with varying patch sizes on a 32-bit model and we evaluate the success rate of each patch size for different quantization levels. 
As shown in Table \ref{tab:experiment6_resnet56}, attack success rates generally decrease as bitwidth is reduced in all patch sizes, with a consistent decline from 32-bit to 2-bit.  
The 10x10 and 12x12 patches exhibit the highest attack success rates, indicating that increasing patch size improves the attack's effectiveness. This supports the hypothesis that larger, more visible patches exert a stronger adversarial influence that is less affected by quantization. 

\begin{table}[ht]
    \centering
    \footnotesize  
    \renewcommand{\arraystretch}{0.9} 
    \setlength{\tabcolsep}{4pt} 
    \begin{tabular}{|c|c|c|c|c|c|}
    \hline
       \textbf{Size} &  \textbf{32-bit} & \textbf{8-bit} & \textbf{5-bit} & \textbf{4-bit} & \textbf{2-bit}\\
    \hline
      6x6   &  86.43  & 83.24  & 79.94 & 76.22 & 73.08\\ \hline
      8x8    & 89.92  & 86.94  & 81.62  & 79.76 & 75.18 \\ \hline
      10x10  & 91.92  & 88.94  & 83.50  & 81.16 & 78.62 \\ \hline
      12x12  & 95.56  & 90.10  & 88.52  & 86.19  & 84.89 \\ 
    \hline
    \end{tabular}
    \caption{ASR (\%) of LAVAN attack against ResNet-56 on CIFAR-10 for different patch sizes.}
    \label{tab:experiment6_resnet56}
\end{table}
\vspace{-3mm}
\section{Key Strategies for Defense Design}
The findings from our study provide valuable insights into the resilience of quantized neural networks to patch-based attacks and offer critical guidance for developing effective defense mechanisms tailored to these specific vulnerabilities. Here, we outline key strategies for defense design, inspired by our experiments:






\noindent\textbf{Enhanced Spatial Awareness:}
Our results indicate that patch-based attacks retain high success rates even when the patch is spatially shifted, suggesting that traditional defenses that rely on detecting exact positions are inadequate. To address this, future defenses could incorporate spatially adaptive filtering techniques that are sensitive to abnormal, high-contrast localized patterns regardless of their exact location. By dynamically analyzing feature maps for unusual activations across various spatial regions, these defenses can better detect adversarial patches that are shifted or scaled.

\noindent\textbf{Gradient-Based Defenses for Localized Features:}
The consistent gradient alignment observed across bit widths for patch-based attacks implies that these attacks retain a stable gradient structure, allowing them to bypass quantization noise. A promising defense strategy could involve regularizing gradient responses to suppress localized, high-gradient regions commonly associated with patch attacks. By applying regularization or gradient smoothing to discourage the model from focusing on compact, high-contrast features, defenses could reduce the impact of patches without sacrificing model performance.

\noindent\textbf{Multi-Bitwidth Training for Robustness:}
Given that patch attacks transfer effectively across different quantization levels, designing defenses that incorporate multi-bitwidth training could improve robustness. By training models to recognize and neutralize patches across a range of bit widths, defenses can increase their adaptability to adversarial signals that exploit quantization inconsistencies. This approach would involve training models with randomly varying bit widths during adversarial training to simulate real-world deployment conditions more accurately.

\noindent\textbf{Cross-Architecture Consistency Checks:}
The high transferability of patch attacks across architectures indicates that a defense mechanism could benefit from comparing activations or gradients across multiple model architectures. Implementing ensemble-based defenses that evaluate inputs across diverse architectures and examine for consistent adversarial patterns could help identify patches that exploit generalized features rather than architecture-specific details. Such cross-architecture ensemble methods would be more resilient to the universal nature of patch-based attacks.

\noindent\textbf{Patch Detection Through Localized Activation Suppression:}
The study reveals that patches induce distinct, high-visibility activations across layers, particularly in early layers. A defense strategy could involve localized activation suppression, where activation values in suspiciously high-contrast regions are reduced to mitigate adversarial impact. This suppression technique could be dynamically activated based on anomaly scores from each spatial region, reducing the patch’s effectiveness without compromising the integrity of the clean input’s features.

\noindent\textbf{Quantization-Aware Defense Training:}
Our results highlight that quantization alone does not neutralize patch attacks effectively. A novel approach could involve Quantization-Aware Defense Training, where models are trained with adversarial patches specifically tailored to simulate quantized conditions. By introducing patches during QAT that are crafted to exploit bitwidth reduction, models can learn to recognize and resist the unique features of quantized patches, leading to improved robustness post-deployment.

\noindent\textbf{Augmenting Training with Diverse Patch Variants:}
Since patch attacks maintain high effectiveness across bitwidths and spatial variations, training models with a diverse set of patch variants—differing in size, position, rotation, and quantization level—could harden defenses against such attacks. By exposing the model to a broad distribution of adversarial patch characteristics, defenses can reduce overfitting to specific configurations, increasing resilience to unforeseen patch variations.

\section{Quantization-Aware Defense Training (QADT)}
Based on our findings, we introduce Quantization-Aware Defense Training (QADT), a strategy that integrates adversarial patch augmentation to improve robustness against patch-based attacks in quantized models. QADT trains models with adversarial patches crafted under simulated quantized conditions, enabling them to adapt to altered gradients and mitigate low-bitwidth vulnerabilities.
Additionally, training with diverse patch variants (e.g., varying in size, position, and rotation) enhances spatial and feature-level awareness, reducing overfitting to specific configurations.

\begin{algorithm}
\caption{Quantization-Aware Defense Training (QADT)}
\label{alg:qraa}
\textbf{Input:} Dataset \(D\), model \(M\), learning rate \(\eta\), number of epochs \(E\), batch size \(B\), adversarial patch generator \(P\), quantization levels \(Q\).\\
\textbf{Output:} Robust quantized model \(M_{QADT}\). \\

1. Initialize model \(M\) with random weights. \\
2. for epoch \(e = 1, \dots, E\):\\
3: \hspace{0.5cm} Shuffle \(D\) and split into batches of size \(B\). \\
4: \hspace{0.5cm} for each batch \(b \in D\): \\
5: \hspace{1.0cm} Extract clean samples \((x, y) \in b\). \\
6: \hspace{1.0cm} Generate diverse adversarial patches \(p\) with generator \(P\) for quantization levels \(Q\): \\
7: \hspace{1.5cm} Vary patch size, position, rotation, and intensity. \\
8: \hspace{1.5cm} Apply simulated quantization on \(p\) using levels \(Q\). \\
9: \hspace{1.0cm} Create adversarial samples \((x_{adv}, y)\) by embedding patches \(p\) into \(x\). \\
10: \hspace{1.0cm} Perform forward pass with \((x, y)\) and \((x_{adv}, y)\): \\
11: \hspace{1.5cm} Compute loss \(L = \text{Loss}(M(x), y) + \text{Loss}(M(x_{adv}), y)\). \\
12: \hspace{1.0cm} Perform backward pass and update \(M\) with \(\eta\). \\
13: \hspace{0.5cm} end for \\
14: end for \\
15: Quantize model \(M\) to levels \(Q\) for deployment. \\
16: return Quantized robust model \(M_{QADT}\). \\

\label{alg:qadt}
\end{algorithm}




The QADT algorithm \ref{alg:qraa} aims to train a robust neural network model against adversarial patch attacks while accounting for model quantization effects. The process begins with the initialization of the model \(M\) with random weights. During training, the algorithm iterates over a specified number of epochs \(E\), shuffling the dataset \(D\) and splitting it into batches of size \(B\) in each epoch. For each batch, clean samples \((x, y)\) are extracted, and a diverse set of adversarial patches \(p\) is generated using the adversarial patch generator \(P\), considering different quantization levels \(Q\). The diversity of the patches is enhanced by varying size, position, rotation, and intensity, followed by simulated quantization to align with the target quantization levels. These adversarial patches are then embedded into the clean samples to produce adversarial inputs \((x_{adv}, y)\). Both clean and adversarial samples undergo a forward pass through the model, with the total loss \(L\) calculated as the sum of the classification losses for both sample types. The model’s weights are updated using backpropagation with a learning rate \(\eta\). This process is repeated for all batches across all epochs. Upon completion of training, the model is quantized to the desired levels \(Q\) to ensure it is optimized for deployment. The final output is a robust, quantized model \(M_{QADT}\) that is resilient to adversarial patch attacks and well-adapted to the challenges of quantized architectures.

As shown in Table \ref{tab:AT}, QADT significantly reduces the attack success rate, achieving 21.3\% ASR (32-bit) and 10.65\% ASR (8-bit), a substantial improvement over standard training (ST) and adversarial training with PGD-7 (AT). 
\vspace{-3mm}
\begin{table}[h]
    \centering
    \footnotesize  
    \renewcommand{\arraystretch}{0.9}  
    \setlength{\tabcolsep}{4pt}  
    \begin{tabular}{|c|c|c|c|c|c|c|}
        \hline
         & \textbf{ST} & \textbf{AT} & \textbf{QADT} & \textbf{ST} & \textbf{AT} & \textbf{QADT } \\ 
         \textbf{Patch Size} & \textbf{32-bit} & \textbf{32-bit} & \textbf{32-bit} & \textbf{8-bit} & \textbf{8-bit} & \textbf{8-bit} \\ \hline
        (10x10) & 87.29 & 44.01 & \textbf{21.3} & 57.51& 39.69 & \textbf{10.65 }\\ \hline
    \end{tabular}
    \caption{ASR (\%) on different trained ResNet20 (CIFAR-10) using the LAVAN attack.}
    \label{tab:AT}
\end{table}
\vspace{-3mm}
\section{Conclusion}
\label{conclusion}

This paper presents a comprehensive study on the robustness and transferability of patch-based adversarial attacks in QNNs. Our findings show that these attacks remain highly effective across various quantization levels and architectures, revealing a systemic vulnerability that quantization alone cannot mitigate. Patch-based attacks achieve high success rates due to their visually prominent, localized nature, which quantization does not effectively disrupt.
Our work emphasizes the need for specialized defenses, such as spatial filtering or localized feature suppression, to protect QNNs from these persistent threats. This study lays the groundwork for future research on robust, targeted adversarial defenses for quantized models.

\section*{Acknowledgment}

This research was partially funded by the Technology Innovation Institute (TII) under the ``CASTLE: Cross-Layer Security for Machine Learning Systems IoT" project.
{
    \small
    \bibliographystyle{ieeenat_fullname}
    \bibliography{main}
}
\clearpage
\setcounter{page}{1}
\maketitlesupplementary

\subsection{Background}
\label{background}
This section provides a detailed overview of key concepts and techniques relevant to the study, including quantization in neural networks, pixel-level and patch-based adversarial attacks, and transferability challenges. It serves as a foundation for understanding the experiments and analyses presented in both the main paper and supplementary material.
\subsubsection{Quantization in Neural Networks}
Deep neural network (DNN) quantization compresses and accelerates models by representing weights, activations, and sometimes gradients with lower bit widths. Quantization is a cornerstone of efficient DNN inference frameworks in industry, enabling deployment on resource-constrained devices such as edge hardware and mobile platforms \cite{jacob2018quantization}. There are two main categories of quantization techniques: Quantization-Aware Training (QAT) \cite{hubara2016binarized, rastegari2016xnor, zhou2016dorefa, li2019additive} and Post-Training Quantization (PTQ) \cite{nagel2020up, li2021brecq, hubara2021accurate, wei2022qdrop}.

\noindent\textbf{Quantization-Aware Training (QAT):} QAT integrates quantization effects into the training process, either by training DNNs from scratch or by fine-tuning pre-trained full-precision models. Given a training dataset $D_{train} = \{x_i,y_i\}^n_{i=1}$, QAT aims to optimize the weights $\omega$ while adapting the model to quantization-induced noise. This approach typically achieves lower quantization loss compared to PTQ but requires access to the full training dataset and additional computational resources for end-to-end training.

The QAT process is often implemented using fake quantization, where weights remain stored in full precision but are rounded to lower bit-width values during inference. The training objective for QAT with fake quantization can be formalized as:

\begin{equation}
    \min_{\omega,\beta} \sum^n_{i=1} L(f(x_i,\omega,\beta), y_i),
\end{equation}

where $L(\cdot)$ is the loss function (e.g., cross-entropy loss), $\omega$ represents the model weights, and $\beta$ denotes the quantization hyperparameter used to map full-precision weights to lower bit-width representations.

\noindent\textbf{Post-Training Quantization (PTQ):} PTQ offers a computationally efficient alternative to QAT, requiring no end-to-end retraining. Instead, PTQ relies on a smaller calibration dataset $D_{cali} = \{x_i,y_i\}^m_{i=1}$, where $m << n$,  to optimize quantization parameters $\beta$. Model weights $\omega$ remain unchanged during PTQ, focusing exclusively on minimizing quantization-induced errors. This process can also be implemented with fake quantization and is formalized as:
\begin{equation}
    \min_{\beta} \sum^n_{j=1} L(f(x_j,\omega,\beta), y_j),
\end{equation}
While PTQ is computationally lightweight, it often incurs higher quantization loss compared to QAT, particularly at lower bit widths.

Recent advancements in PTQ \cite{nagel2020up, li2021brecq, hubara2021accurate, wei2022qdrop} have modeled weight and activation quantization as perturbation problems, using Taylor expansion to analyze loss value changes and reconstruct outputs for individual layers. These methods have significantly improved the accuracy of PTQ models, even under aggressive quantization constraints.

\noindent\textbf{Training Details:}
For CIFAR-10 dataset, all models were trained with a batch size of 128 using Stochastic Gradient Descent (SGD) with a momentum of 0.9 and weight decay set to 
$1\times10^{-4}$. Each model was trained for 50 epochs to ensure adequate convergence across quantization levels and architectures.
\subsubsection{Pixel-Level Adversarial Attacks }
Adversarial attacks involve adding small, carefully crafted perturbations to input data to manipulate a model’s predictions. These perturbations are often imperceptible to humans but are sufficient to mislead the model into making incorrect classifications. Common types of adversarial attacks include:


\noindent \textbf{Fast gradient sign method (FGSM)} \cite{goodfellow2014explaining} is a single-step, gradient-based, attack. An adversarial example is generated by performing a one step gradient update along the direction of the sign of gradient at each pixel as follows:

 \begin{equation}
     x^{adv} = x - \epsilon \cdot sign (\nabla_{x}J(x,y))
 \end{equation}
Where $\nabla J()$ computes the gradient of the loss function $J$ and $\theta$ is the set of model parameters. The $sign()$ denotes the sign function and $\epsilon$ is the perturbation magnitude. 

\noindent \textbf{Projected Gradient Descent (PGD)} \cite{madry2017towards} is an iterative variant of the FGSM where the adversarial example is generated as follows:
 \begin{equation}
x^{t+1} = \mathcal{P}_{\mathcal{S}_x}(x^t + \alpha \cdot sign (\nabla_{x}\mathcal{L}_{\theta}(x^t,y)) )
 \end{equation}
Where $\mathcal{P}_{\mathcal{S}_x}()$ is a projection operator projecting the input into the feasible region $\mathcal{S}_x$ and $\alpha$ is the added noise at each iteration. The PGD attack tries to find the perturbation that maximizes the loss on a particular input while keeping the size of the perturbation smaller than a specified amount. 


In our work, \textbf{pixel-level attacks} specifically refer to \( l_{\infty} \)-PGD attacks with a step size of \( \alpha = \frac{1}{255} \) and a noise budget of \( \epsilon = \frac{8}{255} \). This configuration ensures that the perturbations remain imperceptible while effectively challenging the model's robustness.

\subsubsection{Patch-based Adversarial Attacks} 
Patch-based adversarial attacks introduce highly visible, localized perturbations that redirect the model's focus, often achieving high success rates with minimal modifications \cite{guesmi2023physical}. Unlike pixel-level attacks, which subtly perturb individual pixel values across the entire input, patch-based attacks concentrate their impact within a small, localized region. This focused approach not only makes them more resilient to noise and transformations but also enhances their applicability in real-world, physically realizable scenarios. Patch attacks have been successfully demonstrated across various domains, including image classification \cite{lavan, googleap, li2021generative, chen2022shape} and object detection \cite{guesmi2024dap}, showcasing their high transferability across diverse model architectures.

Despite their demonstrated effectiveness and unique threat potential, the interaction between patch-based attacks and QNNs remains underexplored. Specifically, there is limited research on how quantization impacts the robustness and transferability of patch-based attacks, leaving a significant gap in understanding their efficacy against QNNs deployed in resource-constrained environments.

\textbf{Techniques for Patch-Based Adversarial Attacks:}

\begin{itemize}
    \item \textbf{Localized and Visible Adversarial Noise (LAVAN):} The LAVAN technique \cite{lavan} generates adversarial patches that are both localized and highly visible. These patches are trained iteratively by randomly selecting images and placing the patch at varying locations. This randomization ensures that the patch generalizes well across different contexts and positions, capturing distinctive adversarial features that make it both highly transferable and effective. By applying the patch to diverse scenarios, LAVAN achieves robustness to environmental variations, making it a strong candidate for both digital and physical adversarial settings.
    \item \textbf{Google Adversarial Patch (GAP):} GAP attack \cite{googleap} emphasizes practicality in real-world attacks, addressing challenges associated with traditional \( L_p \)-norm-based adversarial perturbations, such as requiring precise object capture through a camera. GAP creates universal adversarial patches that can be applied to any part of an input, making it more versatile. Furthermore, it incorporates the Expectation over Transformation (EOT) \cite{eot} framework to improve the patch’s robustness against common variations like rotation, scaling, and translation, ensuring its effectiveness under real-world conditions.
    \item \textbf{Deformable Patch Representation (DPR):} DRP attack \cite{chen2022shape} is a type of adversarial attack that focuses on altering the shape and appearance of patches in an image rather than just modifying pixel values. The main idea is to create deformable and dynamic patches that can adapt their form to exploit the vulnerabilities of neural networks more effectively. DPR essentially aims to increase the robustness and transferability of adversarial patches by introducing shape-based deformations, which makes the attack less dependent on specific pixel perturbations and more adaptable across different models and input variations.
\end{itemize}

\subsection{Analyzing the Robustness of Pixel-level based Attacks}
This section evaluates the robustness and transferability of pixel-level adversarial perturbations. The analysis focuses on how these attacks perform across varying quantization levels and model architectures, shedding light on the robustness of QNNs to pixel-level attacks.
\subsubsection{Transferability of Pixel-Level Perturbations Across Quantization Levels}
In this experiment, we evaluate the robustness of different models (ResNet-56, ResNet-20, and VGG-16) under PGD attacks across various quantization levels (Full Precision 32-bit, 8-bit, 5-bit, 4-bit, and 2-bit) by measuring model accuracy at different perturbation magnitudes (\(\epsilon\)). 

\begin{figure}
    \centering
    \includegraphics[width=\linewidth]{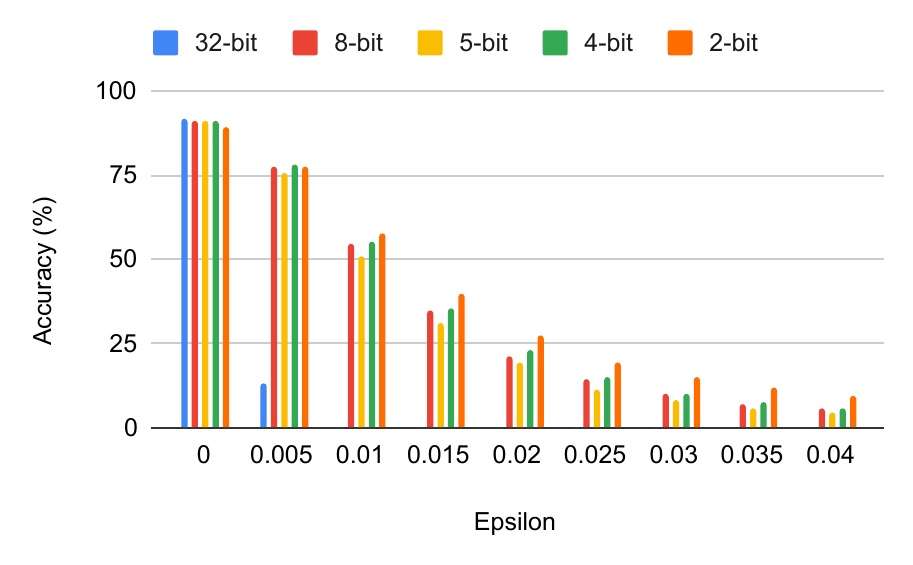}
    \caption{Accuracy of ResNet-56 under PGD attacks across different quantization levels (Full Precision 32-bit, 8-bit, 5-bit, 4-bit, and 2-bit) and perturbation magnitudes ($\epsilon$).}
    \label{fig:pgd_resnet56}
\end{figure}

The results in Figure \ref{fig:pgd_resnet56} indicate a gradual decrease in accuracy as \(\epsilon\) increases for all quantization levels, reflecting the increasing effectiveness of the adversarial attack. At small perturbations (\(\epsilon = 0.005\)), all models maintain relatively high accuracy, indicating minimal adversarial impact. However, as \(\epsilon\) increases, significant differences emerge between the quantized models and the full-precision model. 

Notably, the FP model exhibits the largest drop in accuracy at moderate perturbations (\(0.01 \leq \epsilon \leq 0.02\)), suggesting that quantization provides an inherent robustness against pixel-level adversarial attacks, likely due to gradient misalignment and quantization-induced noise. At intermediate perturbation levels, lower-bitwidth models (e.g., 2-bit and 4-bit) demonstrate slightly better resilience compared to higher-bitwidth quantized models (e.g., 8-bit). This behavior aligns with the hypothesis that the gradient masking effect introduced by the straight-through estimator (STE) in quantized models disrupts the adversarial gradient flow, reducing attack success rates.

As perturbation strength increases further (\(\epsilon = 0.04\)), all models experience a near-complete loss of accuracy, demonstrating that PGD remains effective regardless of quantization at higher perturbation magnitudes. These findings highlight the trade-off between precision and robustness, with quantized models offering moderate resistance to pixel-level adversarial attacks at the expense of clean accuracy, as observed in earlier experiments. Overall, this experiment underscores the potential of quantization as a partial defense mechanism against pixel-level attacks but reinforces the need for additional, quantization-aware adversarial defenses to address higher perturbation levels effectively.

\begin{figure}
    \centering
    \includegraphics[width=\linewidth]{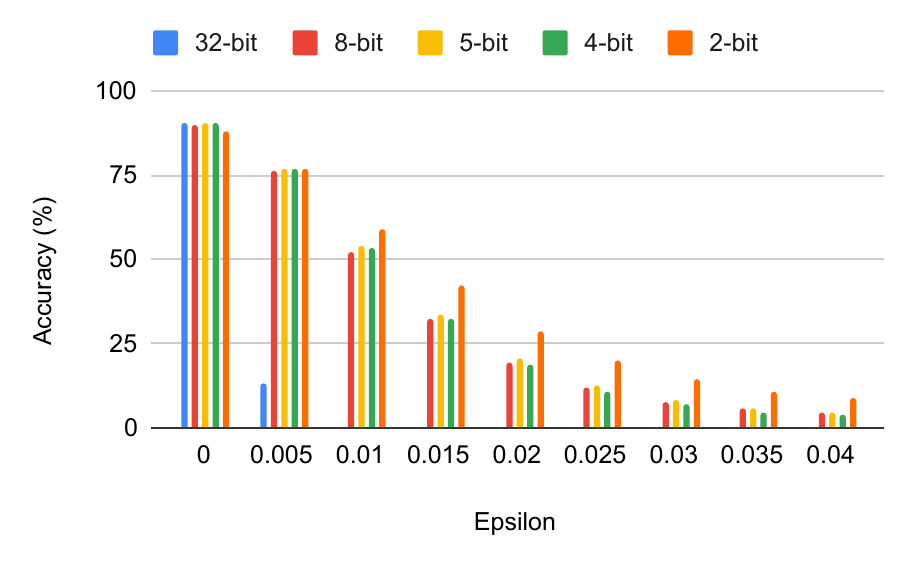}
    \caption{Accuracy of ResNet-20 under PGD attacks across different quantization levels (Full Precision, 8-bit, 5-bit, 4-bit, and 2-bit) and perturbation magnitudes ($\epsilon$).}
    \label{fig:pgd_resnet20}
\end{figure}

The results in Figure \ref{fig:pgd_resnet20} for ResNet-20 under PGD attacks follow the same trend as observed with ResNet-56. Accuracy decreases progressively with increasing perturbation magnitude (\(\epsilon\)) across all quantization levels (Full Precision 32-bit, 8-bit, 5-bit, 4-bit, and 2-bit). At small perturbations (\(\epsilon = 0.005\)), all models retain high accuracy, reflecting minimal adversarial impact. However, at intermediate perturbations (\(0.01 \leq \epsilon \leq 0.02\)), quantized models—particularly lower-bitwidth ones (e.g., 4-bit, 2-bit)—show better robustness than the full-precision model. This trend further supports the hypothesis that quantization disrupts adversarial gradients through gradient misalignment and quantization noise.

At higher perturbation levels (\(\epsilon \geq 0.03\)), accuracy collapses across all models, confirming the effectiveness of PGD at high adversarial strengths, regardless of bitwidth. The overall findings reaffirm that quantization provides partial robustness to pixel-level attacks at moderate perturbation levels, yet all models remain vulnerable under stronger perturbations. 

\begin{figure}
    \centering
    \includegraphics[width=\linewidth]{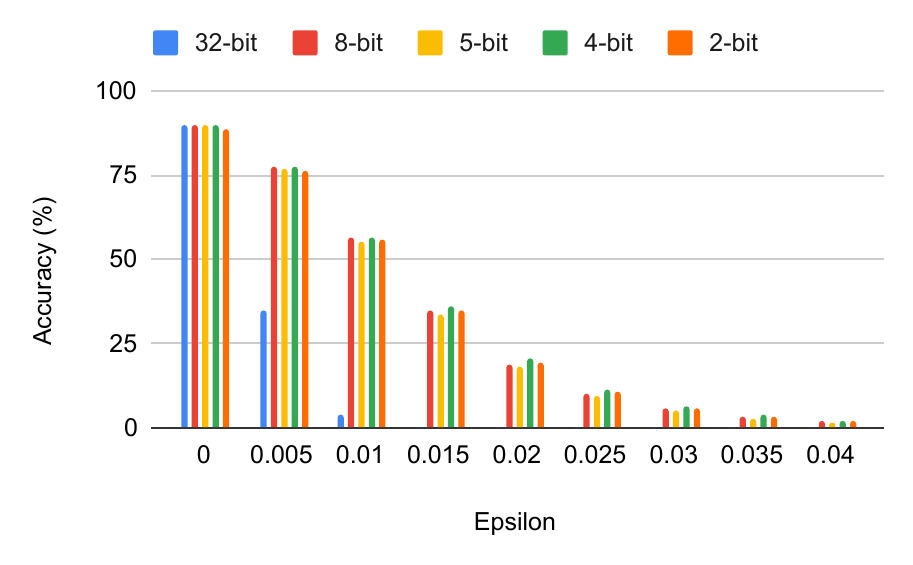}
    \caption{Accuracy of VGG-16 under PGD attacks across different quantization levels (Full Precision (32-bit), 8-bit, 5-bit, 4-bit, and 2-bit) and perturbation magnitudes ($\epsilon$).}
    \label{fig:pgd_vgg16}
\end{figure}
The same trend is observed for the VGG-16 model, where accuracy decreases progressively as the perturbation magnitude (\(\epsilon\)) increases across all quantization levels. Quantized models, particularly those with lower bitwidths, demonstrate slightly better robustness to adversarial perturbations at moderate \(\epsilon\) values compared to the full-precision model, but all models eventually succumb to the attack at higher perturbation levels.
\subsubsection{Transferability of Pixel-Level Perturbations Across Model Architectures}
\begin{figure}
    \centering
    \includegraphics[width=\linewidth]{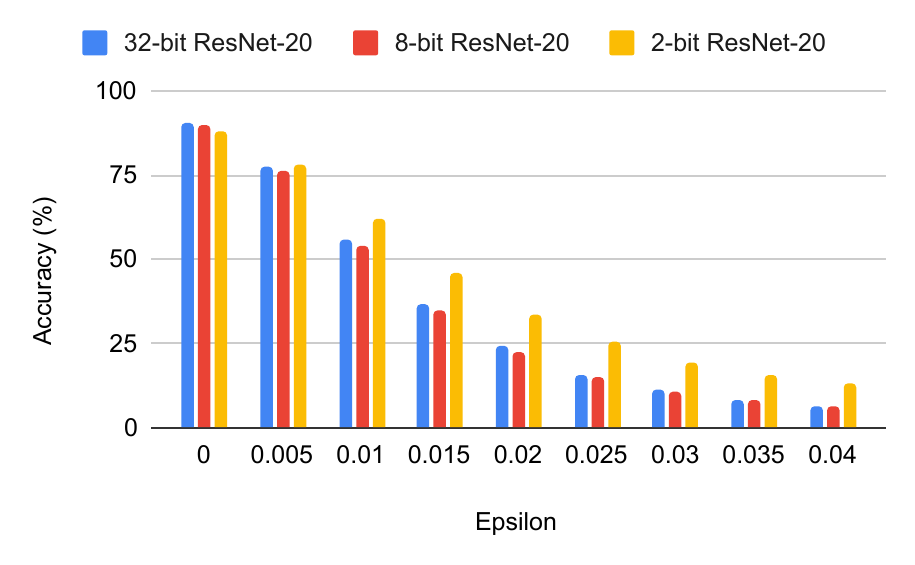}
    \caption{Model accuracy of ResNet-20 under PGD attacks using adversarial examples generated on ResNet-56, evaluated across different quantization levels (32-bit, 8-bit, and 2-bit) and perturbation magnitudes ($\epsilon$).}
    \label{fig:resnet_cross}
\end{figure}

In this section, we report the results for cross-architecture transferability, where adversarial examples generated on ResNet-56 are tested on ResNet-20 under PGD attacks, exhibit a similar trend to the same-architecture experiments, as shown in Figure \ref{fig:resnet_cross}. 

Quantized models exhibit robustness at intermediate \(\epsilon\) values, with the 2-bit model showing better resilience compared to higher-bitwidth models. However, at higher perturbation strengths, accuracy drops significantly across all models. 

\subsection{Analyzing the Robustness of Patch based Attacks}
In this section, we provide additional results on exploring the robustness and transferability of patch-based adversarial attacks on QNNs. Through a series of experiments, we analyze the resilience of patches to various factors, including quantization levels, spatial transformations, patch size, and cross-architecture scenarios, providing a comprehensive understanding of their effectiveness.
\subsubsection{Experiment 1: Feature Map Analysis Across Bit Widths} 

\begin{figure*}
    \centering
    \includegraphics[width=0.9\linewidth]{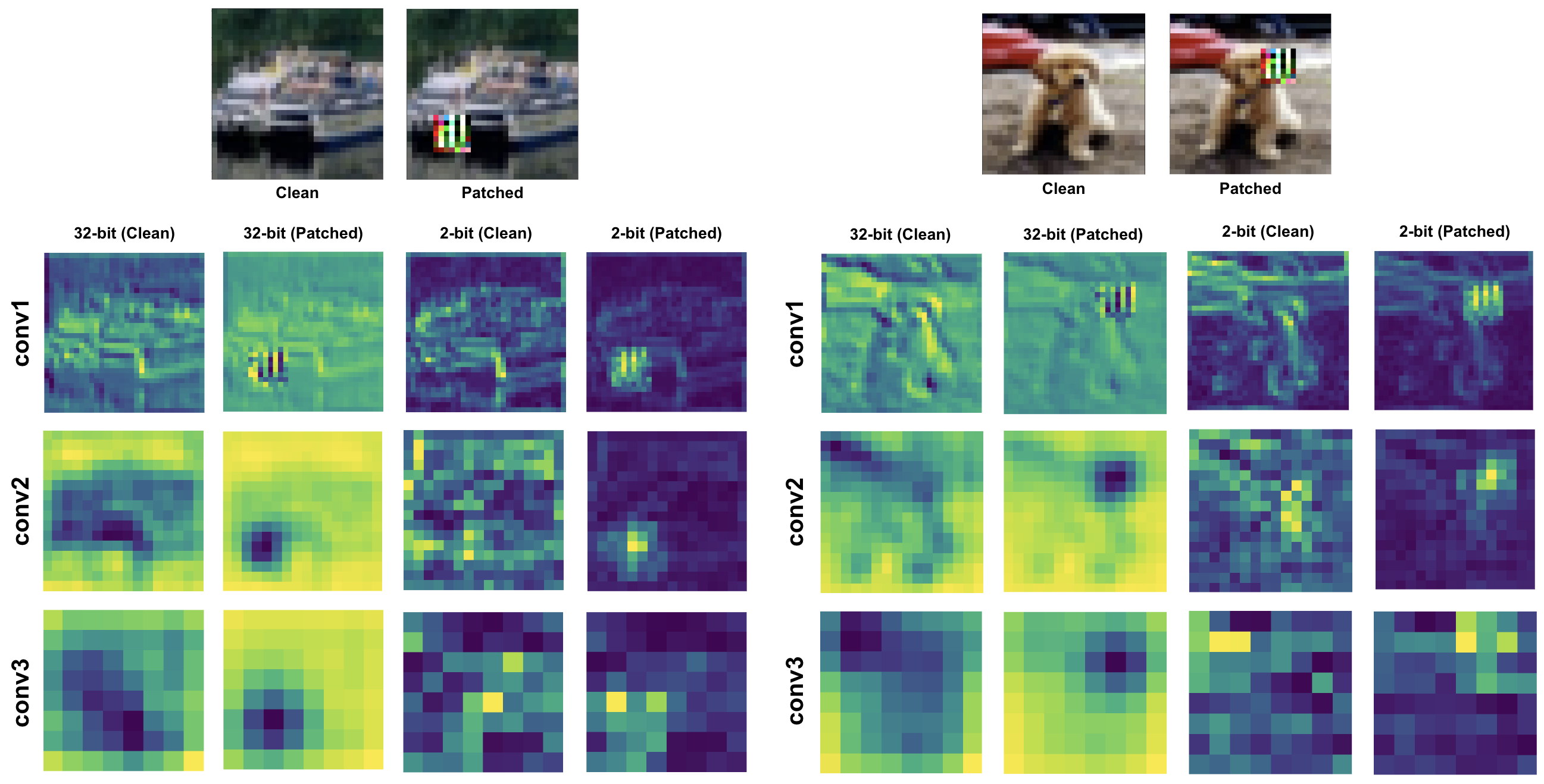}
    \caption{Feature maps of the 32-bit and 2-bit models comparing the clean and patched feature maps for the three first convolutional layers.}
    \label{fig:experiment1_SM}
\end{figure*}

Figure \ref{fig:experiment1_SM} illustrates feature map activations for clean and patched inputs across different quantization levels (32-bit and 2-bit) and convolutional layers. Patch-induced high-visibility activations are evident across layers and persist despite quantization, highlighting the robustness of patches against quantization-induced gradient shifts. Additionally, the results demonstrate that variations in patch location minimally affect the attack’s impact, as the localized perturbations continue to redirect feature activations effectively.

\begin{figure*}
    \centering
    \includegraphics[width=0.9\linewidth]{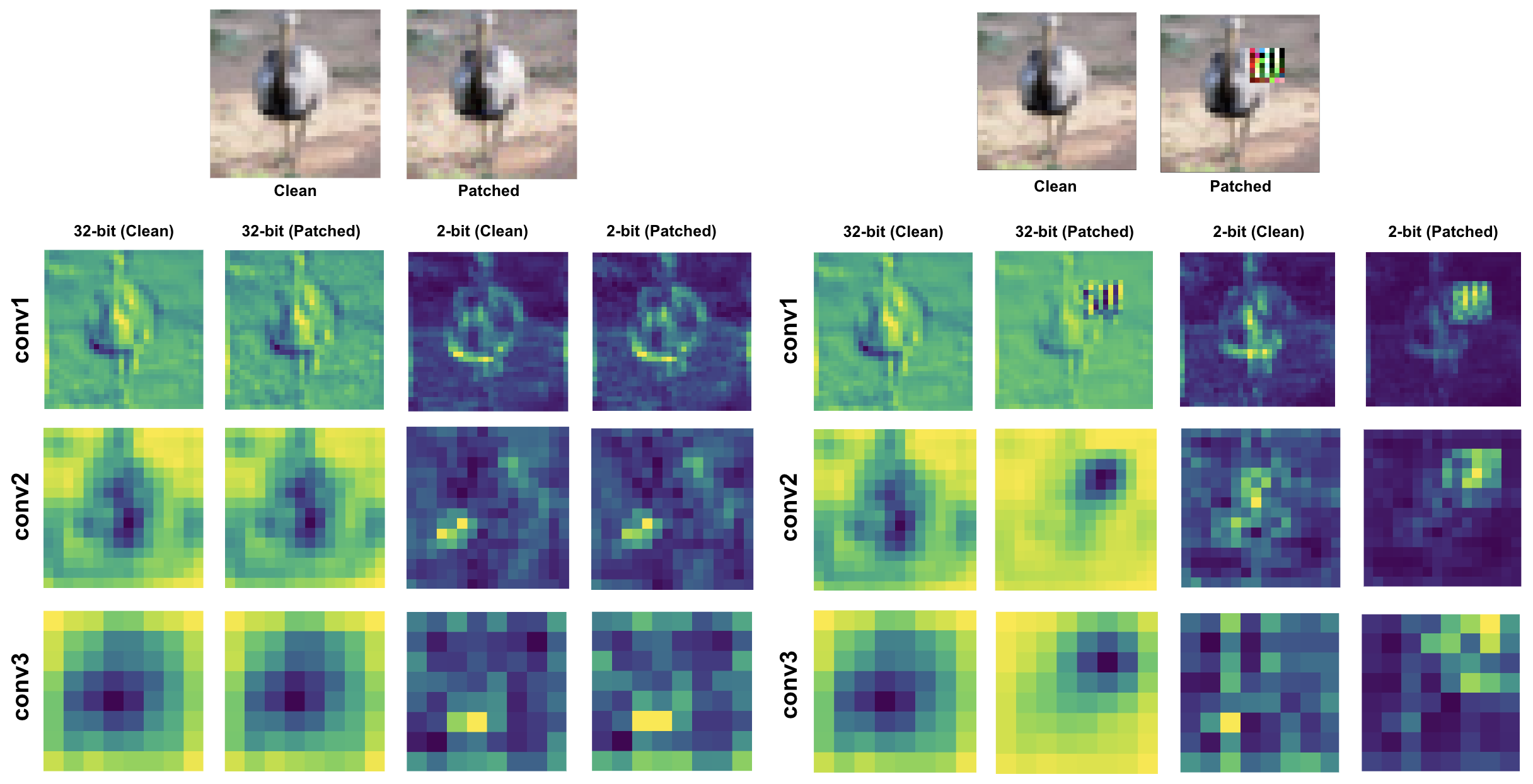}
    \caption{Feature maps of the 32-bit and 2-bit models comparing clean, pixel-level perturbed (left), and patch-based perturbed inputs (right) across the first three convolutional layers.}
    \label{fig:experiment1_pixel}
\end{figure*}

Figure \ref{fig:experiment1_pixel} compare the feature maps of 32-bit and 2-bit models for clean, pixel-level perturbed, and patch-based perturbed inputs. For patch-based attacks (right), there are strong, localized activations in the feature maps, particularly in the regions corresponding to the patch's position. These activations are visible in both the 32-bit and 2-bit models and persist across layers (conv1, conv2, conv3), demonstrating the robustness of patch perturbations even under extreme quantization. 

For pixel-level perturbations (left), the feature maps exhibit subtle, dispersed changes compared to the clean inputs. These changes are more challenging to detect visually, especially in deeper layers like conv2 and conv3. Additionally, the impact of pixel-level perturbations appears less pronounced in the 2-bit model, suggesting that quantization reduces their effectiveness.

In the 2-bit quantized model, patch-based attacks maintain a clear and strong adversarial signal, as indicated by the bright, concentrated regions in the feature maps. This resilience highlights that quantization alone is insufficient to mitigate the impact of patch-based adversarial examples, as their high-visibility features are less affected by the gradient masking and discretization effects introduced by quantization.

In conv1, both pixel-level and patch-based perturbations show the most noticeable changes compared to clean inputs. As the input progresses to deeper layers (conv2 and conv3), the patch-induced activations remain prominent and localized, while the pixel-level perturbations appear to lose their impact or become more diffused.

These results reinforce that pixel-level perturbations are more vulnerable to quantization effects, likely due to gradient misalignment and the tendency of quantization to mask subtle perturbations by rounding them into the same discrete buckets as clean features. In the other hand, Patch-based attacks exploit high-visibility, localized features, making them resilient to quantization. This aligns with their demonstrated effectiveness and high transferability across models with varying architectures and bit widths.


\subsubsection{Experiment 2: Effect of Quantization Levels on Patch Transferability} 
In this experiment, we investigate the transferability of patch-based adversarial attacks across varying quantization levels. The patches used in this evaluation are generated using the Google Adversarial Patch (GAP) technique, which is designed to be robust to spatial transformations and effective across different locations. The attack is targeted, aiming to misclassify all inputs into class 9 on the CIFAR-10 dataset. The results for models with different architectures and quantization bit widths are summarized in Table \ref{tab:exp2_same_architecture}.

The table presents the attack success rates (mean and standard deviation) for models quantized to 32-bit (full precision), 8-bit, 5-bit, 4-bit, and 2-bit. To account for variability due to the random placement of the patch, the attack was executed 10 times, and the mean and standard deviation of the success rates were reported.

Even under this challenging setting—a universal patch generated to target a specific label (class 9) and evaluated across multiple architectures and quantization levels—the patch maintained its effectiveness in fooling the models. The attack achieved high success rates across all architectures in the 32-bit precision setting and remained moderately effective at lower bitwidths (e.g., 8-bit, 5-bit, and even 2-bit). This demonstrates the robustness and efficiency of the patch-based attack in exploiting localized adversarial signals that persist across QNNs.


  
\begin{table}[!htbp]
    \centering
    \small
    \begin{tabular}{|c|c|c|c|c|c|}
    \hline
     \textbf{Model}   & \textbf{32-bit} & \textbf{8-bit} & \textbf{5-bit} & \textbf{4-bit} & \textbf{2-bit}\\
    \hline
      ResNet-20  & 84.71  & 49.61  &  65.31 & 50.45 &  58.31\\ \hline
      ResNet-56  & 84.40  & 56.69  & 65.31  & 54.22  & 47.91 \\ \hline
      VGG-16     &  95.79 & 39.65 & 63.76 & 48.7 & 40.69 \\ \hline
      VGG-19     & 95.71  & 52.04 & 55.03 & 48.90 & 64.24 \\ \hline
  
    \end{tabular}
    \caption{Attack Success Rates (mean (std)) for a targeted patch-based attack on QNNs with varying bitwidths (32, 8, 5, 4, 2-bit) and architectures (ResNet-20, ResNet-56, VGG-16, and VGG-19) on the CIFAR-10 dataset. The attack aims to misclassify all inputs into a specific target class (class 9).}
    \label{tab:exp2_same_architecture}
\end{table}

\begin{table*}
    \centering
    \small
    \begin{tabular}{|c|c|c|c|c|c|c|c|c|c|c|c|}
    \hline
        \textbf{Substitute}  &  & &  & \textbf{Resnet-56} &  & & &  & \textbf{VGG-16} &  & \\
    \hline
       Bitwidth  & & 32-bit & 8-bit & 5-bit & 4-bit & 2-bit& 32-bit & 8-bit & 5-bit & 4-bit & 2-bit\\
    \hline
      Resnet-20 & mean& 67.91  & 40.67  & 45.80 & 39.33 & 43.78 & 40.16  & 35  & 38.30 & 28.29  & 19.74 \\
                &std & 4.3 &  4.57 & 2.22 &  3.92&  2.19&  2.45 &  2.72 & 1.69 &  3 &  3.62\\
    \hline
        \textbf{Substitute}  &  & &  & \textbf{Resnet-56} &  & & &  & \textbf{Resnet-20} &  & \\
    \hline
       Bitwidth  & & 32-bit & 8-bit & 5-bit & 4-bit & 2-bit& 32-bit & 8-bit & 5-bit & 4-bit & 2-bit\\
    \hline
      VGG-16 & mean& 42.9  & 39.08  & 33.18 & 40.67 &  25.74  & 20.21  & 44.85 &  39.86 & 39.9 & 22.16\\
                &std & 2.62 &  3.23 & 4.34 &  3.92 &  2.14 &  3.09 & 2.42&  2.33 & 4.25  &2.54\\
    \hline
        \textbf{Substitute}  &  & &  & \textbf{Resnet-20} &  & & &  & \textbf{VGG-16} &  & \\
    \hline
       Bitwidth  & & 32-bit & 8-bit & 5-bit & 4-bit & 2-bit& 32-bit & 8-bit & 5-bit & 4-bit & 2-bit\\
    \hline
      Resnet-56 & mean& 22.5  & 43.31  & 37.59 & 31.54 & 30.54 & 38.26  & 30.36 & 39.76  &20.36  & 24.57\\
                &std & 2.93 & 2.57  & 3.15 & 3.23  & 2.66  & 2.39  & 3.74 & 2.47  & 2.51  & 2.37 \\
    \hline
        \textbf{Substitute}  &  & &  & \textbf{VGG-16} &  & & &  & \textbf{Resnet-56} &  & \\
    \hline
       Bitwidth  & & 32-bit & 8-bit & 5-bit & 4-bit & 2-bit& 32-bit & 8-bit & 5-bit & 4-bit & 2-bit\\
    \hline
      VGG-19 & mean &  69.12 & 39.76  & 52.84 & 35.76 & 30 & 19.25  &36.99  & 38.6  &43.58 &27.79 \\
             & std & 2.92 & 2.36 & 2.12 & 3.21  & 1.99 & 1.78  & 2.42 &  3.44 & 3.68  &2.82  \\
    \hline
    \end{tabular}
    \caption{Attack Success Rates (mean and std) transfer across QNNs with different bitwidths (32, 8, 5, 4, 2-bit) and architectures (Resnet-20, Resnet-56, and VGG-16) on CIFAR-10 for a targeted attack. }
    \label{tab:cross_architecture_SM}
\end{table*}

\subsubsection{Experiment 5: Transferability of Patch Across Architectures} 

We further evaluate the transferability of targeted GAP patches across different architectures. The success rate results are summarized in Table \ref{tab:cross_architecture_SM}, highlighting the patch's ability to maintain its adversarial effectiveness when applied to models with varying structures.

Patches generated on one model architecture (e.g., Resnet-20, VGG-16, Resnet-56) maintain a considerable success rate when transferred to other architectures, indicating that the attacks retain effectiveness across architectural differences.
For example, patches crafted on Resnet-20 show high attack success on Resnet-56 across bitwidths, with success rates remaining above 40\% in many cases, even at lower bitwidths.

Although success rates tend to decrease as bitwidths are reduced, the decrease is generally moderate rather than dramatic. For instance, when transferring patches from Resnet-20 to VGG-16, the success rate at 32-bit is 40.16\% and decreases to 19.74\% at 2-bit. Patches generated on VGG-16 transferred to Resnet-20 show success rates from 44.85\% at 8-bit to 22.16\% at 2-bit, suggesting that quantization weakens but does not eliminate attack success.

Certain architectures appear to retain higher success rates across different bitwidths. For instance, patches generated on Resnet-56 achieve 43.31\% on Resnet-20 at 8-bit, showing relatively high cross-architecture transferability.
Similarly, patches from VGG-16 transferred to Resnet-56 achieve moderate success rates, with 40.67\% success at 4-bit, indicating that the VGG-based patch patterns are robust across Resnet architectures.

The high cross-architecture transferability, with patches generated on one model retaining effectiveness on another, suggests that these patches exploit generalized vulnerabilities within QNNs that are not highly specific to architecture. This indicates that quantized models share common feature representations that patches can disrupt, regardless of architectural nuances.



\subsubsection{Experiment 6: Impact of Patch Size }  

\begin{table}
    \centering
    \small
    \begin{tabular}{|c|c|c|c|c|c|}
    \hline
       \textbf{Size} &  \textbf{32-bit} & \textbf{8-bit} & \textbf{5-bit} & \textbf{4-bit} & \textbf{2-bit}\\
    \hline
     6x6     &88.17& 76.29 & 77.59 & 88.20 &90.95\\ \hline
      8x8    &  88.76  & 72.16  &  78.74 & 81.46 & 86.42 \\ \hline
      10x10  &  88.17 & 76.29  & 77.59  & 88.20 & 90.95 \\ \hline
      12x12  & 91.12  & 73.71  & 79.31  & 79.21  & 83.54 \\
    \hline
    \end{tabular}
    \caption{Attack Success Rates (\%) for different patch sizes on VGG-19.}
    \label{tab:experiment6_vgg19}
\end{table}

In this experiment, we analyze the effect of patch size on the attack success rates of LAVAN-based adversarial patches. By varying the patch size on the VGG-19 architecture, we aim to evaluate how the size of the adversarial region impacts the patch's ability to fool quantized QNNs with varying bit widths. The results are reported in Table \ref{tab:experiment6_vgg19}.

Attack success rates are high across all patch sizes and bitwidths, with most success rates above 70\% and reaching as high as 91.12\% at 32-bit and 12x12 patch size.


Attack success rates generally remain stable across bitwidths, with only slight reductions in some configurations. For example, the 8x8 patch decreases from 88.76\% at 32-bit to 72.16\% at 8-bit but then rises to 86.42\% at 2-bit.
Similarly, other patch sizes exhibit only minor variations in success rates across bitwidths, indicating that quantization has minimal impact on patch effectiveness. The 2-bit quantized model shows surprisingly high success rates for all patch sizes, suggesting that even extreme quantization does not significantly reduce vulnerability to these patches.






\subsubsection{Impact of Traditional Pre-Processing Based Defenses}

\begin{figure}
    \centering
    \includegraphics[width=\linewidth]{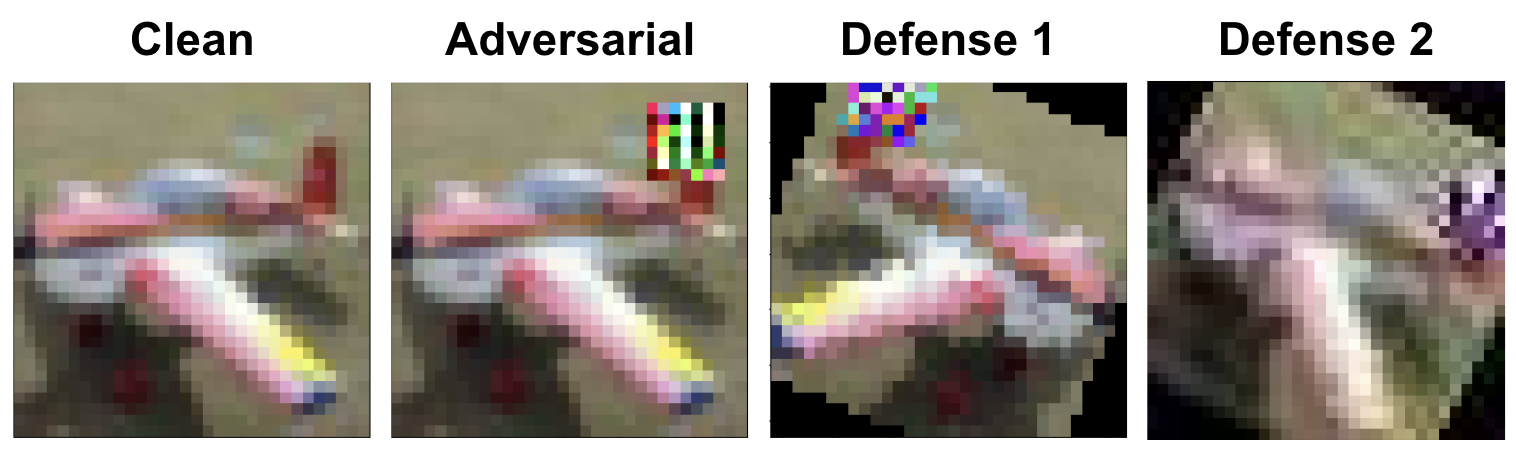}
    \caption{Illustration of a clean sample, an adversarial sample, and the corresponding outputs after applying Defense 1 and Defense 2.}
    \label{fig:experiment7}
\end{figure}

This experiment evaluates the robustness of traditional pre-processing-based defenses against patch-based adversarial attacks on QNNs across various bit widths (32-bit, 8-bit, 4-bit, and 2-bit).
Two sets of input pre-processing based defenses were tested (see Figure \ref{fig:experiment7}): \textbf{Defense 1}, which applies random cropping, rotation (\(\pm20^\circ\)), horizontal flips (50\% probability), and Gaussian noise (\(\mu = 0, \sigma = 0.1\)), and \textbf{Defense 2}, which incorporates all transformations from Defense 1 with an additional JPEG compression step (quality levels 50–70). The results reveal that clean accuracy is consistently high without defenses (\(\sim74.45\%\) across bit widths), while adversarial accuracy drops significantly, with success rates as low as \(19.61\%\) for 8-bit models under attack. 

As illustrated in Table \ref{tab:experiment7}, Defense 1 demonstrates moderate improvements in robustness against adversarial patches, achieving \(25.46\%\) robust accuracy for 32-bit models and \(21.30\%\) for 8-bit, while maintaining relatively high clean accuracy (\(71.51\%\) for 32-bit). Defense 2 provides slightly better robustness for adversarial inputs in some cases (e.g., \(24.75\%\) for 32-bit and \(16.08\%\) for 8-bit) but incurs a more significant drop in clean accuracy (\(64.97\%\) for 32-bit and \(62.43\%\) for 8-bit). Both defenses exhibit diminished effectiveness at lower bit widths (e.g., 2-bit), where adversarial accuracy remains above \(24.79\%\) even with Defense 2, highlighting the limited ability of these techniques to mitigate adversarial patch effects. 
The results reveal that quantization does not inherently enhance robustness against patch-based attacks. Even with defenses in place, low-bitwidth models (e.g., 2-bit QNNs) remain vulnerable.
\begin{table}
    \centering
    \small
    \begin{tabular}{|c|c|c|c|c|c|c|}
    \hline
        & \textbf{Input} & &\textbf{32-bit} & \textbf{8-bit}  & \textbf{4-bit} & \textbf{2-bit}\\
    \hline
        No defense & Clean & mean & 74.45 & 74.2  &  75.4 & 74.45\\ \hline
        No defense & Adv. & mean & 25.73 & 19.61 & 20.32  & 25.72 \\ \hline
        Defense 1 & Clean & mean & 71.51 & 70.67 &  71.78 & 71.79 \\ \hline
                  &   & std & 0.60  & 0.66  & 0.39  & 0.78 \\ \hline
        Defense 1 & Adv. & mean & 25.46 & 21.30 &  25.63 & 25.31 \\ \hline
                  &   & std & 0.69  & 1.35  &  1.20 & 1.45 \\ \hline
        Defense 2 & Clean & mean & 64.97 &  62.43 & 65.37 & 64.83 \\ \hline
                  &  & std &  0.52 & 0.71  & 0.52  & 0.69 \\ \hline
        Defense 2 & Adv. & mean & 24.75 & 16.08 & 19.21  & 24.79\\ \hline
                  &  & std & 1.98 & 1.32 & 1.53  & 1.61 \\
    \hline
    \end{tabular}
    \caption{Robust Accuracy (\%) for different input pre-processing techniques.}
    \label{tab:experiment7}
\end{table}

\end{document}